\documentclass[10pt]{article}
\usepackage[letterpaper,margin=0.82in]{geometry}
\usepackage[T1]{fontenc}
\usepackage[utf8]{inputenc}
\usepackage{newtxtext,newtxmath}
\usepackage{microtype}
\usepackage{graphicx}
\usepackage{booktabs}
\usepackage{tabularx}
\usepackage{threeparttable}
\usepackage{amsmath}
\usepackage{xcolor}
\usepackage{hyperref}
\usepackage[numbers,sort&compress]{natbib}
\usepackage{caption}
\usepackage{placeins}
\usepackage{float}
\usepackage{listings}
\definecolor{linkblue}{HTML}{245B78}
\hypersetup{colorlinks=true,allcolors=black,pdfauthor={Arash Ahmadi, Dingjing Shi, Yaser M. Banad},pdftitle={Adaptive Ecological Momentary Assessment with a Hybrid Language Model: Formative Expert Review and Retrospective Evaluation}}
\newcommand{\system}{\textsc{EMA-E4B}}

\title{\textbf{Adaptive Ecological Momentary Assessment with a Hybrid Language Model: Formative Expert Review and Retrospective Evaluation}}
\author{Arash Ahmadi$^{1}$ \qquad Dingjing Shi$^{2}$ \qquad Yaser M. Banad$^{1,*}$\\[8pt]
\small $^{1}$School of Electrical and Computer Engineering, University of Oklahoma\\
\small Norman, OK 73019, USA\\[4pt]
\small $^{2}$School of Psychology, Georgia Institute of Technology\\
\small Atlanta, GA 30332, USA\\[5pt]
\small $^{*}$Correspondence: Yaser M. Banad, \href{mailto:bana@ou.edu}{bana@ou.edu}}
\date{}

\begin{document}
\hyphenpenalty=10000
\exhyphenpenalty=10000
\maketitle

\begin{abstract}
Ecological momentary assessment (EMA) measures experience in daily life, but fixed questionnaires and schedules collect information of uneven value and can interrupt participants. We present and retrospectively evaluate \system{}, a hybrid framework for question selection and prompt timing. Separate ridge models propose an item set and delay; a supervised Gemma 4 E4B language layer produces the final structured response and explanation. Evaluation distinguishes proxy action performance, output conformity, and formative judgments of response quality. The data contain 4,372 records from 79 participants and yield 3,516 sequential cases under a participant separated split. One involved domain expert preferred the complete hybrid response in 15 of 20 decisive comparisons, with two ties among 22 reviews. On 75 reused development cases, hybrid question utility and timing similarity were 0.832 and 0.818; the head alone reached 0.852 and 0.818. A separate 60 case comparison with untouched E4B under the same head gave action differences of $-0.0031$ and $-0.0105$. Thus, the language layer produced structured responses with action scores comparable to or slightly below the reference configurations, while the expert feedback favored the complete hybrid response. These observations establish a concrete, inspectable framework and clarify the distinct roles of action scoring and response review. Repeated adaptive administration and practical effects on measurement and participant burden remain future research.
\end{abstract}

\section{Introduction}

Mental health states change over hours and days, yet many research and clinical instruments summarize experience over much longer recall periods. Intensive longitudinal measurement can expose fluctuations that a single retrospective score cannot represent \citep{shiffman2008ecological,onnela2016harnessing}. Ecological momentary assessment addresses this gap through repeated reports made near the time and setting of an experience \citep{shiffman2008ecological}. The method has become common in psychology, behavioral medicine, and digital phenotyping because it supports the study of person specific dynamics in natural settings \citep{wrzus2023ecological,onnela2016harnessing}.

Repeated measurement also creates a design problem. More prompts and more items can improve temporal resolution, but each request competes for attention and may increase interruption, fatigue, or missingness \citep{williams2021compliance,wrzus2023ecological}. Published EMA protocols vary markedly in length and schedule, and compliance reporting is not uniform \citep{williams2021compliance}. A fixed protocol cannot use the fact that some constructs are stable for one person while other constructs are changing at the present moment \citep{li2024ask,schneider2024just}.

Prior adaptive methods address parts of this problem. Computerized adaptive testing and information gain can shorten a survey while retaining information about an estimated state \citep{schneider2024just,li2024ask}. Context aware prompting can time requests near activity transitions, and just in time adaptive intervention research formalizes decisions that depend on changing vulnerability and receptivity \citep{aminikhanghahi2019context,nahum2016just,klasnja2015microrandomized}. These methods offer clear statistical targets, but each usually optimizes one layer of the problem, such as survey length, current state classification, context sensitive delivery, or intervention choice \citep{schneider2024just,li2024ask,aminikhanghahi2019context}.

Language models introduce a different design space. Transformer architectures support attention based sequence modeling, and large pretrained models can perform tasks from textual demonstrations \citep{vaswani2017attention,brown2020language}. Instruction tuning and preference learning further adapt model responses to task instructions and human preferences \citep{ouyang2022training,rafailov2023direct}. These capabilities motivate a language interface for structured actions and explanations; they do not guarantee that an emitted action is correct. Mental health applications require targeted evaluation rather than reliance on fluent language alone \citep{kjell2024beyond,guo2024large}.

Model choice also matters. Large hosted models may offer strong general reasoning, but they raise cost, reproducibility, privacy, and governance questions for repeated sensitive data processing \citep{bommasani2021opportunities,onnela2016harnessing}. Small open weight models provide a path to local inference and task adaptation, yet their lower capacity can make structured reliability and narrow domain evaluation more important \citep{team2026gemma,van2025survey}. Results from one language model therefore cannot be assumed to transfer to another model of a similar nominal size \citep{team2026gemma,van2025survey}.

This study asks whether a hybrid small model system can support two linked EMA decisions: which questions should be asked next, and when the next prompt should occur. The deterministic head proposes the bounded action from recent observations, personal baselines, trends, variability, and prompt position. The language layer produces the final action and a concise explanation for review. The observed next response is hidden from inference and later supplies retrospective outcome information.

The paper presents a concrete hybrid framework and an evaluation that separates proxy action performance, output conformity, and formative response quality. This separation is the central contribution: action scores describe the emitted decisions under explicit retrospective rules, whereas expert review assesses the complete response. The language layer did not improve the measured action endpoints, although one involved expert preferred the complete hybrid in most decisive comparisons. Completed head only and same scaffold comparisons are central to this interpretation; supplementary analyses document external models and subsequent preference experiments.

\section{Related Work}

\subsection{EMA, intensive longitudinal data, and burden}

EMA reduces retrospective recall demands and permits study of change at the time scale of lived experience \citep{shiffman2008ecological}. Meta analytic work reports high mean compliance but substantial variation in protocol design, samples, incentives, and reporting \citep{williams2021compliance,wrzus2023ecological}. These summaries do not imply that a shorter protocol always improves compliance. They instead motivate designs that measure information retained as burden changes.

The source data were collected for a study of mental health and social contact during the first COVID 19 lockdown in the Netherlands \citep{fried2022mental}. Speyer, Murray, and Kievit later used the same item structure to illustrate a dynamic structural equation model of momentary depression, loneliness, social media use, and their interaction \citep{speyer2024investigating}. Our work does not reproduce that model. It uses the sequential EMA structure to train and evaluate an action policy.

\subsection{Adaptive question selection and prompt timing}

JITA EMA adapts item administration through psychometric information and a stopping rule \citep{schneider2024just}. Ask Less, Learn More estimates the information supplied by question answers and adapts survey length \citep{li2024ask}. SEP EMA detects activity transitions to select less disruptive prompt times \citep{aminikhanghahi2019context}. Adaptive data collection methods also model adherence while balancing measurement quality and burden \citep{monacelli2023adaptive}. These studies motivate our two action formulation. Our difference is the combined selection of heterogeneous EMA items and a prompt delay, an explicit explanation for both decisions, later outcome feedback, and a paired expert review interface.

Just in time adaptive intervention research treats intervention choice as a sequence of decisions that depend on changing vulnerability and receptivity \citep{nahum2016just}. Micro randomized trials estimate the near term effects of such decisions and make the decision points, availability rules, and proximal outcomes explicit \citep{klasnja2015microrandomized}. Our present analysis is not a micro randomized trial and does not estimate a causal effect. It borrows the discipline of a stated decision point, bounded action space, and later outcome assessment.

\subsection{Language model adaptation and evaluation}

Instruction tuning uses demonstrations and human preferences to improve task following \citep{ouyang2022training}. LoRA reduces the number of trained parameters by learning low rank updates \citep{hu2021lora}. Group relative policy optimization supports sampled policy improvement without a separate value model \citep{shao2024deepseekmath}. Direct preference optimization learns from chosen and rejected responses through a classification style objective \citep{rafailov2023direct}. Earlier research runs used these ideas, but the evaluated system in this paper uses an outcome fit ridge head and completion only supervised training. It does not use DPO.

LLM as a judge can scale qualitative comparison, but judge bias, position effects, and disagreement with humans must be measured \citep{zheng2023judging}. We use forward and reversed response order and report order consistency. We do not treat an LLM judge as a substitute for the expert's review.

Mental health language model studies also require a distinction between plausible language and supported action. Published reviews describe uneven evidence, risks of overstatement, and the need for human oversight \citep{kjell2024beyond,guo2024large}. This motivates our separation of a rule based language constraint, an expert safety rating, and a model judge safety rating rather than calling all three measures safety.

\begin{table*}[t]
\centering
\caption{Position of the present study relative to closely related adaptive measurement methods. Entries describe reported method components and targets, not clinical validation.}
\label{tab:related}
\small
\begin{tabularx}{\textwidth}{lXXXXX}
\toprule
Method & Adaptive item set & Adaptive prompt time & Optimization or evaluation target & Natural language rationale & Expert pairwise review \\
\midrule
JITA EMA \citep{schneider2024just} & Yes & Not evaluated & Momentary state classification & No & No \\
Ask Less, Learn More \citep{li2024ask} & Yes & No & Information reconstruction & No & No \\
SEP EMA \citep{aminikhanghahi2019context} & No & Yes & Response rate & No & No \\
Adaptive adherence design \citep{monacelli2023adaptive} & Limited & Yes & Data quality and burden & No & No \\
\system{} & Yes & Yes & Retrospective proxy utility & Yes & Yes \\
\bottomrule
\end{tabularx}
\end{table*}

\section{Problem Formulation}

At decision time $t$, the policy receives only observations that would be available at that time. Let $H_{i,t}$ denote the observed history for participant $i$. The system returns an item subset $S_{i,t}$, a delay $d_{i,t}$, a priority label, and a short rationale. Training cases use $|S_{i,t}|\in\{4,5,6,7,8\}$, while every development and test case fixes $|S_{i,t}|=6$. The system produces both decisions in one response, but fits question valuation and timing urgency separately. It does not optimize a joint objective over measurement value and cumulative participant burden. The delay is selected from $\{30,60,180,240,720\}$ minutes.

\begin{figure}[H]
\centering
\includegraphics[width=\textwidth]{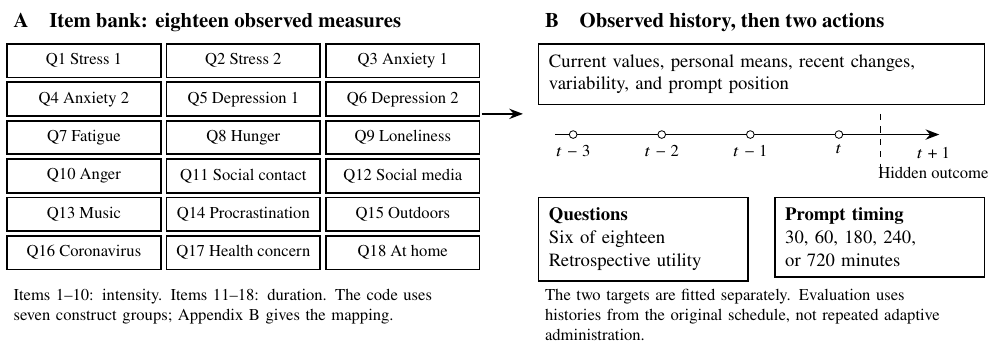}
\caption{Observed item context and the two action outputs. Human readable labels summarize the source instrument. The later scheduled response remains hidden from inference. No repeated adaptive trajectory is simulated.}
\label{fig:cases}
\end{figure}

The next response $Y_{i,t+1}$ is hidden from the prompt. After an action is produced, this response supports an outcome utility score. This separation prevents direct leakage of the value later used to judge the action. Figure~\ref{fig:cases} shows the data path.

\section{Method}

\subsection{Data source and case construction}

The authorized source file contains 4,372 scheduled EMA records from 79 pseudonymized participants. Of these records, 3,918 contain all 18 item values, 30 contain some but not all item values, and 424 contain no item values. The original study recruited 80 Leiden University students and administered four daily prompts from March 16 to 29, 2020 \citep{fried2022mental}. One participant is absent from the cleaned file, with no documented reason. A nominal 79 participants by 14 days by four prompts would give 4,424 slots, which is 52 more than the recorded rows. The source file does not establish the reason for those absent slots; they are distinct from present rows with missing responses. Each prompt contains 18 ordinal items about the past three hours. Items 1 to 10 measure intensity from 1 to 5. Items 11 to 18 measure duration using five ordered categories.

We generated a decision case only when both the current and immediately next scheduled row contained both depression items, loneliness, and social media use, and at least two prior scheduled rows existed. The builder does not skip a missing next row to find a later completed response. Each case includes up to three prior scheduled rows, current values, person specific means through the current response, recent endpoint changes, variability, and prompt position. Prior rows may contain missing items; person means omit missing item values. All history features stop at time $t$. Item correlations used by the diversity adjustment were estimated from training cases. The full set contains 3,516 cases. Deterministic participant hashing creates 2,640 training cases from 60 participants, 346 development cases from eight participants, and 530 test cases from 11 participants. A participant appears in one split only. Requiring a later observed response conditions every outcome analysis on prompt completion. The current data do not identify unobserved outcomes after a missed prompt, so they cannot estimate how a policy would affect missingness.

Data use followed the split map rather than a single untouched test analysis. The action head used all 2,640 training cases; these cases also formed the language adapter's training pool. The 346 development cases supplied the 150 case expert inventory, the 22 completed expert reviews, both judge analyses, and the fixed 75 case benchmark. A 60 case sample from seven development participants supported the same scaffold language layer diagnostic. The policy head v2 experiment used 470 test cases after a separate 60 case exclusion recorded by that experiment. The subsequent analyses are exploratory rather than fresh validation.

\subsection{Observable action head}

The action head consists of two ridge models fit on the 2,640 training cases. The question model uses current value, person mean, signed and absolute deviation, recent standard deviation, recent endpoint change, trend indicators, item index, and construct indicators. Its ridge penalty is 0.1. The timing model uses current risk rank, current elevation above the person mean, prompt position, recent endpoint change summaries, variability summaries, quadratic terms, and three interactions. Its ridge penalty is 1.0. The descriptive risk tier is high when current depression mean or loneliness is at least 3, moderate when either is at least 2, and low otherwise. This rule is not a clinical risk assessment. Every person mean uses data only through time $t$. Hidden future values, outcome scores, and target risk labels are excluded from inference features.

The question target for item $j$ is
\begin{equation}
a_j=.42\delta_j+.23b_j+.20c_j+.10f_j+.05z_j+.02,
\end{equation}
where $\delta_j=\min(|Y_{i,t+1,j}-Y_{i,t,j}|/2,1)$, $b_j=\min(|Y_{i,t+1,j}-\bar{Y}_{i,t,j}|/2,1)$, $c_j$ is the normalized need score for the item's construct, $f_j$ is a fixed focal item weight, and $z_j$ marks a changing recent trend. The construct need score adds absolute change in the construct mean, 0.5 times its positive next response elevation above the personal construct mean, and 0.25 if any item in that construct has an increasing recent trend. This sum is rounded to three decimals. It is divided by the largest construct score for that case, with the denominator floored at $10^{-6}$ and the ratio capped at one, to obtain $c_j$. Focal weights are 1.0 for both depression items and loneliness, 0.8 for social media use, 0.45 for both anxiety items, and 0.35 for social contact and procrastination. Other items receive zero focal weight.

For a selected set $S$, the retrospective set value is
\begin{equation}
V(S)=\max\!\left\{10^{-6},\sum_{j\in S}a_j+.045\,C(S)-\sum_{\substack{j,k\in S\\j<k}}p_{jk}\right\},
\end{equation}
where $C(S)$ is the number of represented constructs. If the absolute training set correlation of two items is at least 0.45, $p_{jk}=.22|\rho_{jk}|\min(a_j,a_k)$. A same construct pair below that threshold receives $p_{jk}=.05\min(a_j,a_k)$; other pairs receive zero. Each unordered pair is counted once. The reference $S_g$ is built through greedy marginal selection under the same budget. We report question utility $Q(S)=\min(V(S)/V(S_g),1)$, which ranges from zero to one and is higher when more of the greedy reference value is retained. The reference is greedy rather than a global combinatorial optimum. Greedy ties favor the lower item index. The coefficients, focal weights, and urgency thresholds are researcher specified; the expert's pairwise choices did not estimate them.

The hidden timing target begins with
\begin{equation}
h=\min\{1,.25r_t+.30r_{t+1}+.20\Delta r+.20\min(\Delta_{\max}/2,1)+.15\min(b/1.5,1)\},
\end{equation}
where risk ranks low, moderate, and high as zero, one, and two; $\Delta r$ is a nonnegative risk increase; $\Delta_{\max}$ is the largest absolute next change in depression, loneliness, or social media use; and $b$ is current positive elevation above the personal depression or loneliness mean. The rule maps $h\geq.75$ to 30 minutes, $h\geq.60$ to 60, $h\geq.38$ to 180, $h\geq.20$ to 240, and lower values to 720. For chosen delay $d$ and rule target $d^*$, timing similarity is
\begin{equation}
T(d)=\exp\{-0.72|\log(d/d^*)|\}.
\end{equation}
This value also ranges from zero to one, and higher is better. The coefficients multiply unnormalized ranks: persistent moderate risk contributes 0.55 before other terms, and persistent high risk contributes 1.10 before clipping, so the latter always maps to the 30 minute target. The hidden urgency is rounded to three decimals before threshold mapping during scoring. The observed interprompt gap is recorded but does not enter this rule; overnight and daytime changes are not normalized by elapsed time. The delay is a designer specified rule applied to a later response. It is not an observed optimal schedule or a causal effect estimate.

The head predicts $a_j$ and $h$ from observable features, then greedily selects items and applies the fixed delay map. A missing next item is carried forward from its current value for retrospective utility. A missing current item uses a zero sentinel in the ridge features. Item predictions are not clipped before greedy selection; predicted urgency is clipped to $[0,1]$ before thresholding and rounded to four decimals only for the returned field. In a deterministic audit of all 63,288 item predictions for the 3,516 frozen cases, values ranged from 0.05645 to 0.96325, with no negative values. This observation does not constrain future inputs. Appendix~\ref{app:implementation} specifies the feature scaling, missing values, and correlation calculation.

\subsection{E4B language layer}

The language layer is based on the instruction tuned Gemma 4 E4B model \citep{team2026gemma}. It receives the observable case context and the action head recommendation. The prompt asks it to preserve the action fields, but this is an instruction rather than a programmatic constraint. The evaluator scores the emitted action. On the 75 case benchmark, the model changed the proposed question set in 37 cases and preserved every proposed delay. It returns one JSON object with selected questions, delay, timing label, priority, priority construct, expected information retention, personalization basis, expert review note, and a brief rationale.

We trained a rank 16 LoRA adapter through 100 completion only supervised optimizer steps with a learning rate of $2\times10^{-5}$. LoRA alpha was 16, dropout was zero, and the target modules were the attention and feed forward projections. The run used a batch size of one, eight accumulation steps, a maximum sequence length of 3,072, a five percent warmup, a cosine schedule, seed 3407, and bfloat16 precision. The launch used one visible GPU and an effective batch of eight. The recorded epoch fraction of 0.3030303 corresponds to 800 example presentations from the 2,640 case pool, not complete exposure to that pool. The implementation pads individual examples without packing and delegates sampling to the Trainer; no custom sampler is configured and the exact encountered case order was not retained. Supervised targets contain the action head output and an explanation derived only from observable features. They do not expose the later response to E4B. The training loss was 0.4907. No reward optimization or DPO changed the evaluated checkpoint. The loss is an optimization diagnostic, not evidence of policy benefit. The artifact records the base repository name but not an immutable base revision, which limits exact reproduction.

\begin{table*}[!htbp]
\centering
\caption{Component ownership, objective, and evidence scope.}
\label{tab:components}
\small
\begin{tabularx}{\textwidth}{lXXX}
\toprule
Component & Fit or use & Data & Role in the reported system \\
\midrule
Ridge action head & Regression to targets derived from the hidden next response & 2,640 training cases & Proposes questions and delay \\
E4B LoRA & Completion only supervised learning to head generated targets & Pool of 2,640; 800 presentations & Produces the final action and explanation \\
DPO holdout adapter & Pairwise preference learning & 15 expert pairs; five participant held out pairs & Exploratory preference fit \\
GPT 5.6 Luna & Two order rubric application; no model fitting & 75 development cases per comparison & Secondary model judge \\
\bottomrule
\end{tabularx}
\end{table*}

\subsection{Forecast response and comparison configurations}
Policy and forecast outputs use separate prompts through the same loaded checkpoint for each condition. The hybrid uses E4B with its merged supervised adapter for both; the original condition uses instruction E4B without that adapter. The forecast prompt receives the observed history but not the head recommendation or generated action. It asks for next depression mean (the mean of Q5 and Q6), loneliness (Q9), social media use (Q12), a descriptive risk tier, rationale, and uncertainty note. The horizon is the next recorded scheduled observation, not the delay selected by the policy. The 100 step supervised fit used policy completions, not forecast targets. Expert comparisons include both outputs. This paper reports forecast wrapper conformity and expert response ratings, not numerical forecast accuracy. Appendix~\ref{app:implementation} gives a schema illustration.

\begin{table}[!htbp]
\centering\small
\caption{Distinct model contrasts. Head entries refer only to the policy prompt; forecast prompts receive no head proposal. SFT denotes the supervised adapter.}
\label{tab:configurations}
\begin{tabularx}{\textwidth}{lXXXX}
\toprule
Comparison & Candidate / comparator & Head proposal & Adapter & Prompt evidence\\
\midrule
Same scaffold & Hybrid / instruction E4B & Both & SFT / none & Evaluation configuration\\
Expert review & Hybrid / instruction E4B & Hybrid only & SFT / none & Saved detailed prompts, version 2\\
Later DPO judge & DPO / prior hybrid & Both & DPO / SFT & Saved generation metadata, version 3\\
External benchmark & Hybrid / direct models & Hybrid only & SFT / none & Exact cache match; external script\\
\bottomrule
\end{tabularx}
\end{table}
The expert responses were generated with detailed rationale prompts and a 768 token cap. The older 200 case E4B inventory used a 220 token cap, while external models used 320. Thus, expert and benchmark outputs are not interchangeable even when they describe the same participant moment. The older inventory does not retain its prompt version; its launch script and generation metadata establish the model and head configuration. The later DPO judge evaluates a different response contrast and cannot be used as a concordance test against the expert's hybrid versus instruction E4B choices.

\subsection{Output checks}

Earlier reinforcement learning explored forecast and action rewards, but it did not optimize the supervised checkpoint evaluated here. Appendix~\ref{app:implementation} distinguishes those experiments from the reported hybrid.

The requested output contract specifies one answer wrapped JSON object, the exact item budget, a delay from the five specified choices with a matching timing label, and no hidden reasoning block. The frozen language screen checks diagnostic, treatment plan, guaranteed, definitely, and must ask expressions, with exceptions for specified negations of diagnosis. Its structural delay check accepts 15 to 1,440 minutes, which is broader than the five choices in the prompt. Consequently, a passing screen does not establish full adherence to that requested action space. We call the resulting binary measure the research language and action constraint rate. It is a lexical and structural screen rather than a clinical safety assessment. Expert safety ratings and judge safety ratings are separate measures.

\subsection{Expert preference family}

One domain expert reviewer who is also a study coauthor reviewed paired outputs in a formative portal study. The reviewer knew the study aims and helped define the review dimensions. The available review record does not establish response identity masking or randomized display order, and the retained portal interface identifies the training conditions. We therefore do not claim a blinded expert comparison. Each record showed observable EMA history, the later observed response, and both candidate outputs. The reviewer selected the hybrid response, the original E4B response, or a tie. Only the hybrid received seven absolute ratings from 1 to 5 for forecast, question selection, timing, rationale, burden, safety, and overall quality. The anchors were 1 unacceptable, 3 revise, and 5 ready. These ratings are not paired differences. We call this the evaluated hybrid; the expert feedback informed its continued research use.

The expert evidence is frozen at the August 6, 2026 snapshot: 22 reviews drawn from eight source participant histories. We summarize decisive preferences and retain ties. The Wilson interval treats judgments as individual observations; the exploratory cluster interval resamples source participants. Written comments remain attached to their cases. These reviews informed selection of the hybrid for the research portal and later preference experiments, so they are formative model selection evidence rather than independent validation. The later inventory contains 150 E4B comparisons for continued review, divided into 75 priority and 75 extension cases. The frozen review snapshot and current inventory are different selections. Appendix~\ref{app:implementation} describes their provenance. The explanation question asked whether reasoning was clear and supported by the information shown; it did not provide separate factual or action consistency ratings.

\subsection{Secondary model judge}

GPT 5.6 Luna applied the seven dimension rubric to each pair in both response orders through the Codex SDK. A subsequent DPO versus supervised hybrid comparison and eight external model comparisons each used 75 development cases. We retain these judgments as a supplementary assessment of complete response quality, not a substitute for expert review. Appendix~\ref{app:judge} gives the service settings, blinding procedure, aggregation rule, and uncertainty estimates.

\begin{figure}[H]
\centering
\includegraphics[width=\textwidth]{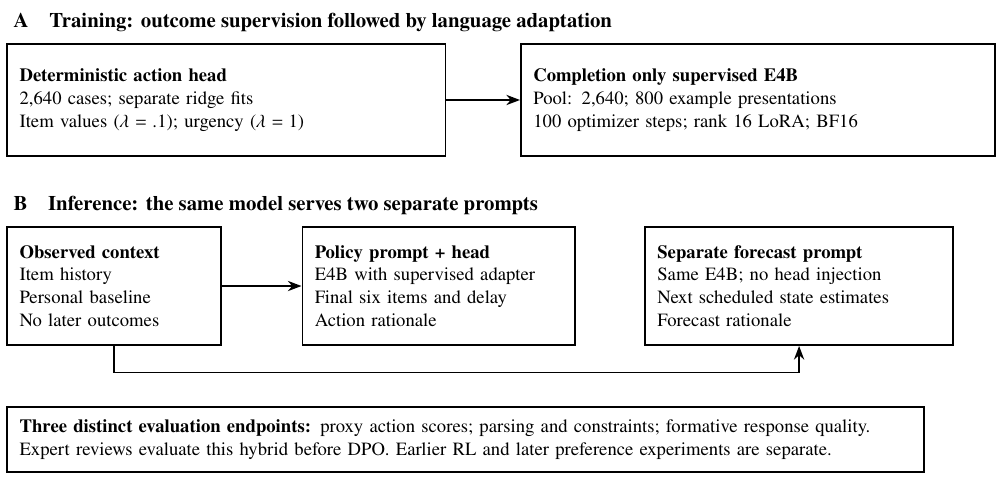}
\caption{The hybrid's two training stages and separate policy and forecast prompts. Only the policy prompt receives the head recommendation. The forecast targets the next scheduled observation. The evaluated hybrid is the supervised version before DPO.}
\label{fig:system}
\end{figure}

Figure~\ref{fig:system} summarizes the evaluated system.

\section{Evaluation Protocol}

\subsection{Outcome comparisons}

The evaluated hybrid was compared with untouched E4B under the same action scaffold on a 60 case sample from seven development participants. The diagnostic takes zero based development rows $0,5,\ldots,295$ in frozen file order. Stored paired intervals use 5,000 case resamples rather than participant clusters. A later policy head v2 experiment used 470 test cases after a separate 60 case exclusion and 10,000 case resamples. A five fold participant out of fold analysis supplied a second descriptive estimate. The v2 experiment compares a gradient boosting question ranker and ExtraTrees timing model with the ridge head; no language model is involved. These analyses are exploratory because the resampling unit is too narrow and the cases have informed research decisions. Question utility is the primary action endpoint. Timing similarity is secondary because its target is a designer rule rather than an observed optimal delay.

\subsection{Fixed prompt small model benchmark}

We froze 75 development split cases from six participant histories before inspecting external model results. Cluster sizes are 16, 15, 15, 15, 9, and 5 cases. The same cases had already supported expert review preparation, so the benchmark is descriptive and cannot serve as a new confirmatory test. The comparison contains the evaluated hybrid, original E4B without an injected head recommendation, and eight pinned open instruction checkpoints from Gemma, Qwen, Phi, SmolLM, and Granite families.

Every external model received the same observable prompt, action budget, forecast request, and output contract, but it did not receive the fitted head output or the hidden timing rule. Inference used bfloat16, deterministic greedy decoding, and at most 320 new tokens. Each checkpoint passed a one case load test and a five case prefix before the full resumable run. We report policy parse, exact budget, research language and action constraint rate, question utility, timing similarity, exact forecast wrapper conformance, paired generation latency, and peak allocated GPU memory. Malformed policy responses receive zero action utility when the required fields cannot be recovered. Recovered action fields can retain utility even when another contract check fails. The question reference uses the recovered subset size; the exact budget check is separate. In the ten recorded benchmark runs, every budget failure had an empty recovered subset and received zero question utility. Forecast wrapper failures often contain fenced JSON or a reasoning preamble, so this endpoint measures strict contract conformance rather than forecasting ability. The run did not preserve cap hit rates or a lenient parse sensitivity analysis. Participant cluster percentile ranges use 5,000 draws. Participants are sampled with replacement, and all their cases are retained for each sampled occurrence. Means are case weighted, so a participant with more cases contributes more weight. Paired differences are formed before resampling, preserving the candidate and comparator pairing. The endpoints are the 2.5th and 97.5th percentiles; benchmark intervals linearly interpolate between ordered draws, whereas judge intervals use the nearest ordered draw. Expert ranges use 20,000 draws and interpolated percentiles. The expert preference implementation divides by at least one decisive judgment, so a draw with no decisive judgments would contribute zero. They are descriptive sensitivity summaries rather than confidence intervals because the cohort contains only six clusters. The E4B responses exactly match the stored 200 case generation inventories. Their old metadata do not record a prompt version, although the launch script documents head injection for the hybrid only. Their latency and memory are unavailable.

\subsection{Expert and judge comparisons}

The expert analysis uses the latest saved review per case. All 22 cases had one saved submission, so no overwritten verdict was available for an intrarater estimate. The decisive preference rate excludes ties from the denominator. We report a Wilson interval for the 20 decisive judgments and a descriptive source participant resampling range. Rating ranges also resample source participants. Each LLM judge analysis contains 75 cases judged in two orders, which yields 75 paired scores rather than 150 independent observations. Its participant resampling ranges are descriptive. No adjustment is applied for the eight model judge comparisons.

\begin{table*}[!htbp]
\centering
\caption{Analysis cohorts and reuse. Evaluation participants are separate from the original action head and E4B training participants. Later analyses reuse development cases.}
\label{tab:cohorts}
\small
\begin{tabularx}{\textwidth}{lrrrX}
\toprule
Cohort & Cases & Source participants & Split & Reuse and overlap \\
\midrule
Action head fit / E4B pool & 2,640 & 60 & Training & Model fitting \\
Development pool & 346 & 8 & Development & Source for expert and judge studies \\
Same scaffold diagnostic & 60 & 7 & Development & Earlier model decision \\
Policy head v2 & 470 & 11 & Test & Separate 60 case exclusion \\
Completed expert review & 22 & 8 & Development & All 22 also occur in the E4B judge set \\
DPO versus hybrid judge & 75 & 8 & Development & Same contexts, different comparison from expert \\
External benchmark and judge & 75 & 6 & Development & Seven cases overlap completed expert reviews \\
\bottomrule
\end{tabularx}
\end{table*}

\subsection{Project selection criteria}

The recorded project selection criteria require empirical structured validity, exact budget, and research language and action constraint rates of at least 0.99. Both gate versions required a question utility gain of at least 0.02 and a positive timing similarity gain. The scaffold diagnostic required interval lower limits above $-0.02$ for question utility and $-0.03$ for timing similarity. The later policy head v2 gate required nonnegative interval lower limits. These are distinct recorded criteria, not a common retrospective test. The 0.02 question threshold was a project decision rather than a clinically derived minimum effect. It was not changed after the reported comparisons. These empirical selection criteria were fixed for the respective comparisons, not preregistered for the whole research program. A sample pass rate does not establish a future reliability guarantee.

\section{Results}

\subsection{Structured reliability and outcome evidence}

All 60 evaluated diagnostic responses passed the specified parsing, item budget, and research language checks. Against untouched E4B under the same action scaffold, hybrid minus reference question utility was $-0.0031$ (case bootstrap interval $[-0.0109,0.0026]$), and timing similarity was $-0.0105$ ($[-0.0316,0]$). Higher is better for both measures, so the point estimates slightly favor untouched E4B. All 60 delays also belonged to the five requested choices. The observed differences did not reach the project's predefined improvement thresholds.

\subsection{Policy baseline evidence}

The head only baseline reached question utility of 0.852 and timing similarity of 0.818 on the 75 case descriptive cohort. The hybrid reached 0.832 and 0.818. Hybrid minus head question utility was $-0.020$, with a descriptive participant resampling range of $[-0.041,0.002]$. The hybrid preserved the head delay in every case but preserved the head question set in 38 of 75 cases, or 50.7\%. It preserved the exact item order in 22 cases. These action scores primarily reflect the fitted head.

An exact mean over all $\binom{18}{6}$ question subsets gave random subset utility of 0.580. Hybrid minus random utility was 0.252, with a descriptive range of 0.211 to 0.276. A constant 720 minute delay reached timing similarity of 0.780. Hybrid minus constant timing was 0.038, with a descriptive range of $-0.026$ to 0.166. The small timing difference and its wide range show that the timing endpoint is dominated by the rule's frequent 720 minute target.

\begin{table}[!htbp]
\centering
\caption{Simple policy baselines on the reused 75 case development cohort.}
\label{tab:baselines}
\small
\begin{tabular}{lcc}
\toprule
Policy & Question utility & Timing similarity \\
\midrule
Ridge head only & \textbf{0.852} & \textbf{0.818} \\
\system{} hybrid & 0.832 & \textbf{0.818} \\
Exact random six item subset & 0.580 & not applicable \\
Constant 720 minute delay & not applicable & 0.780 \\
\bottomrule
\end{tabular}
\end{table}

The hybrid selected 720 minutes in 56 of 75 cases, 240 minutes in six, 60 minutes in ten, and 30 minutes in three. It never selected 180 minutes. The 13 choices below 180 minutes overlap the instrument's three hour recall window. The project rule labeled 62 cases low, ten moderate, and three high. These are algorithmic tiers, not validated clinical categories. All three high tier cases received 30 minutes and retained a depression item. Ten moderate cases received 60 minutes; seven retained a depression item and four retained loneliness. This pattern shows burden concentration in the cases labeled more urgent. It does not establish that the added prompts are useful or acceptable.

\FloatBarrier

\subsection{Fixed prompt benchmark evidence}

\begin{figure}[!htbp]
\centering
\includegraphics[width=\textwidth]{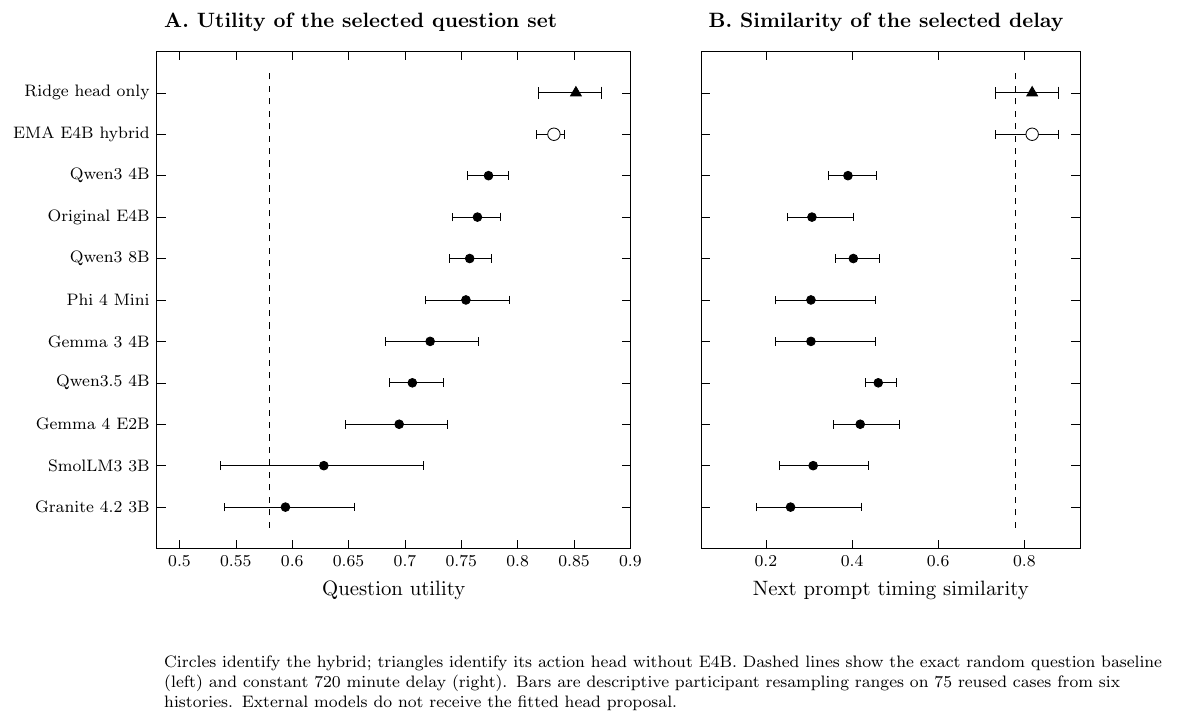}
\caption{Question utility and next prompt timing similarity for the hybrid, its ridge head alone, original E4B without a head recommendation, and eight external instruction models. All estimates use the same 75 development cases. Bars show descriptive participant resampling ranges from six histories. Dashed lines identify simple policy baselines; Table~\ref{tab:baselines} gives their values.}
\label{fig:benchmark-utility}
\end{figure}

The evaluated hybrid had the highest question utility and timing similarity in the ten system comparison, but the head only baseline above had higher question utility and equal timing similarity. Hybrid question utility was 0.832, compared with 0.774 for Qwen3 4B Instruct. The paired difference was 0.058, with a descriptive participant resampling range of 0.033 to 0.080. Timing similarity was 0.818, compared with 0.461 for Qwen3.5 4B. The paired difference was 0.357, with a descriptive range of 0.228 to 0.438. Figure~\ref{fig:benchmark-utility} reports all system estimates. These values compare a task fitted hybrid that receives a head recommendation with direct instruction models that do not. They do not rank the base architectures or isolate E4B.

Contract reliability differed sharply among models. All ten systems returned parseable policy actions in most cases, yet research language constraint rates ranged from zero for Gemma 3 4B to 1.0 for \system{} and Qwen3.5 4B. The zero indicates that every response triggered at least one frozen lexical or structural rule. It is not evidence that every response was clinically unsafe. Only four systems satisfied the exact forecast wrapper in every case. Appendix Figure~\ref{fig:benchmark-contract} separates these dimensions because policy parsing alone would hide contract failures.

\FloatBarrier

\subsection{Expert preference evidence}

The reviewer selected \system{} in 15 cases, original E4B in five cases, and a tie in two cases. The decisive hybrid preference was 75\%. Its Wilson 95\% interval was 53.1\% to 88.8\%. A descriptive participant resampling range was 57.1\% to 94.1\%. This was an interim, formative review of 22 cases. The later 150 case inventory supports continued expert review; it is not 150 completed ratings. No stopping or alpha spending rule was registered. The review informed the decision to retain \system{} for expert access, so later review of the same inventory remains exploratory.

Mean absolute ratings for \system{} were 4.09 for forecast, 3.86 for question selection, 4.05 for timing, 4.09 for rationale, 3.82 for burden, 3.86 for expert safety, and 4.05 overall. The comparator did not receive these ratings. Burden ratings were one score of 2, four of 3, 15 of 4, and two of 5. Expert safety ratings were eight scores of 3, nine of 4, and five of 5. No safety rating was below 3. The two written comments show why a single global preference is not sufficient: one favored the hybrid timing but the original explanation, and one asked for more frequent engagement checks despite a good hybrid prediction.

\begin{figure}[!htbp]
\centering
\includegraphics[width=.96\textwidth]{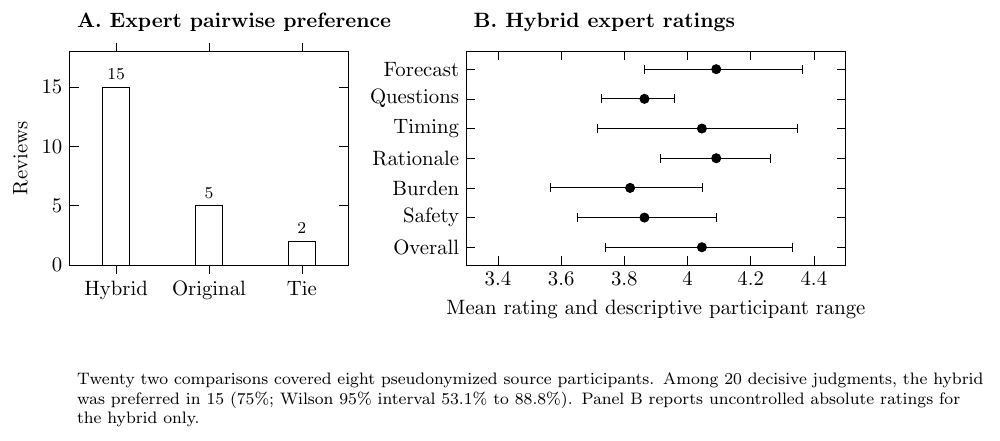}
\caption{Formative expert review of \system{}. Panel A shows all pairwise choices. Panel B shows uncontrolled absolute ratings for the hybrid response with descriptive participant resampling ranges.}
\label{fig:expert}
\end{figure}

\subsection{Supplementary comparisons}
The external model judge favored the hybrid over each direct instruction model, with agreement gated win scores from 0.693 to 0.980. These results supplement the action and expert endpoints. Appendix~\ref{app:judge} reports this analysis and a distinct later comparison of a DPO adapter with the supervised hybrid. The latter does not compare the same responses as the expert's review and is not an estimate of agreement with the expert's preferences.

\begin{table}[!htbp]
\centering
\caption{Current evidence for \system{} and related experiments.}
\label{tab:results}
\small
\begin{threeparttable}
\begin{tabularx}{\textwidth}{lXl}
\toprule
Evidence source & Result & Interpretation \\
\midrule
Development diagnostic validity & Parse, budget, language constraint = 1.00 & Specified checks passed \\
Same scaffold comparison & Question utility $-0.0031$; timing similarity $-0.0105$ & Below improvement thresholds \\
Head only baseline & Question 0.852; timing 0.818 & Higher question; equal timing \\
Observable policy v2 test & Question gain $+0.0100$; timing gain $-0.0017$ & Question gain below gate; timing unresolved \\
Expert preference & 15 hybrid, 5 original, 2 tie & Formative preference for hybrid \\
External model judge & Win scores 0.693 to 0.980 & Hybrid favored; reused cohort \\
DPO holdout & No action change & Explanation effects unmeasured \\
\bottomrule
\end{tabularx}
\begin{tablenotes}\footnotesize
\item Effects come from different case sets and answer different questions. They must not be pooled.
\end{tablenotes}
\end{threeparttable}
\end{table}

\FloatBarrier

\section{Discussion}
\system{} demonstrates the feasibility of combining an observable action head with a language layer that produces structured recommendations and case specific explanations. The formative expert review favored the complete hybrid response in 15 of 20 decisive comparisons. The head only configuration achieved slightly higher question utility and identical timing similarity, so the present action scores primarily reflect the fitted head. The completed same scaffold comparison provides a complementary result: supervised E4B produced action scores close to those of untouched E4B given the same recommendation.

The framework makes these endpoints inspectable rather than interchangeable. Proxy utility measures the selected set under a frozen retrospective rule. Timing similarity measures agreement with a rule for delay. Parsing and lexical screens characterize response conformity. Expert preference describes the perceived quality of the complete forecast, action, and explanation. A favorable response review can coexist with unchanged or slightly lower action scores because the reviewer sees more than the action alone.

The external benchmark compares a task adapted hybrid with direct instruction models that do not receive its head proposal. It therefore assesses configured systems, not intrinsic model quality. Supplementary automated judgments describe another assessment of those responses; they do not replace expert review. The separate DPO experiment reports unchanged actions on its small held out sample, while effects on explanation quality were not evaluated. These completed comparisons identify useful next questions without changing the model evaluated in the formative study.

\section{Limitations and Future Work}
\paragraph{Formative evidence.} The reviewer is one involved domain expert and study coauthor. The 22 reviews provide initial feedback rather than independent validation, and response identity masking is not established. Continued review of the 150 case inventory can extend this evidence. Independent reviewers and another participant cohort are future extensions; the present report does not depend on obtaining them.

\paragraph{Observed histories and repeated administration.} The retrospective decisions use histories collected under the original full item schedule. Repeated six item selection would leave some past values unobserved. The present analysis does not specify a validated strategy for maintaining state estimates under that observation pattern. Whether short surveys supplement full assessments, replace them, or follow a full item initialization period remains undecided. Sequential adaptive replay and explicit state maintenance are future evaluations, not completed experiments.

\paragraph{Proxy endpoints and burden.} Question valuation and timing urgency are fitted separately. Fewer items per prompt do not imply lower cumulative burden, especially when prompts are frequent. The source protocol contains no outcomes under the proposed schedule, and 30 or 60 minute actions overlap the three hour recall window. Prospective research will assess burden, response completion, measurement value, recall overlap, and prompt limits.

\paragraph{Exploratory sampling.} Participant groups are separated for fitting and evaluation, but development reuse and subsequent test analyses make the comparisons exploratory. The benchmark contains six participant histories from a historical Dutch student cohort \citep{fried2022mental}. Participant resampling and omission analyses describe sensitivity to these histories, not an untouched replication.

\paragraph{Response assessment.} The expert explanation rubric asked about clarity and support from the shown information; it did not separately measure factual accuracy, action consistency, coverage of timing and question selection, uncertainty, or reviewer error detection. A targeted factual audit and explanation specific comparisons will address these properties. Token budget and lenient parsing sensitivity will clarify external model performance under other output conditions. The system remains a research tool, not a diagnostic, crisis triage, or treatment system.

\section{Reproducibility and Governance}

Case construction, reward computation, model configuration, and portal artifacts are retained in the local project record. The expert review cache preserves 150 E4B comparisons and the 22 earlier reviews. Participant identifiers are hashed, so the records are pseudonymized rather than anonymous. Model inference prompts exclude the hidden next response, but expert and judge review prompts include it as an evaluation reference. The research portal exposes only \system{}; DPO and 12B candidates remain research only. An internal evidence ledger identifies the artifact for each reported number.

The project record does not include a secondary analysis ethics determination or a provider retention attestation for the hosted judge. Confirmation of applicable data permissions and institutional requirements remains an unresolved author action before public submission. The preprint does not release participant trajectories or raw hosted judge inputs.

We recommend preregistering the external cohort protocol before collection. The registration should freeze item mappings, case construction, prompt template, model checkpoint, decoding settings, endpoint definitions, exclusion rules, missing data handling, cluster unit, bootstrap seeds, and selection thresholds.

\section{Conclusion}
We present a hybrid framework for EMA question selection and prompt timing, with a deterministic proposal head and a language layer that produces the final structured response. Its evaluation separates proxy action scores, output conformity, and formative response quality. All 60 diagnostic responses passed the specified checks, and one involved expert preferred the complete hybrid in 15 of 20 decisive comparisons. The head only and same scaffold results indicate that the language layer's measured action scores were comparable to or slightly below the reference configurations. These results support a transparent first implementation and motivate future evaluation of explanation usefulness and repeated adaptive administration.

\section*{Acknowledgments}

Domain expertise informed the methodological design and evaluation priorities. Arash Ahmadi designed and implemented the training and evaluation pipeline. Dr. Yaser M. Banad supervised the research direction and governance plan.

\section*{Author Contributions}

Arash Ahmadi: conceptualization, methodology, software, analysis, visualization, and writing. Dingjing Shi: domain methodology. Yaser M. Banad: supervision and research direction.

\section*{Competing Interests and Funding}

The sole formative expert reviewer is a study coauthor. This research received no external funding.

\section*{Data and Code Availability}

The source study materials are available from the original authors' OSF repository \citep{fried2022mental}. The cleaned working file, participant level outputs, and trained model weights are not publicly released with this preprint. The manuscript source includes the equations, implementation details, and aggregate results needed to inspect the reported comparisons. A separate source snapshot for author review contains selected methods and scoring code with file hashes; it is not a public repository or a complete computational reproduction package. Full computational reproduction requires the original case data and model artifacts; public access to those artifacts has not yet been established.

\appendix
\section{Supplementary Model Judge Results}\label{app:judge}

A subsequent all 20 pair DPO adapter was compared with the prior supervised hybrid, with a candidate win score of 0.520 and a descriptive participant resampling range of 0.486 to 0.556. Order consistency was 0.747. The overall DPO minus supervised hybrid rating difference was 0.020, with a descriptive range of $-0.057$ to 0.103. The judge contexts include the 22 expert cases, but the response contrast differs: The reviewer compared the supervised hybrid with original instruction E4B, while this judge compared DPO with the supervised hybrid. We therefore do not report expert agreement or Cohen's $\kappa$ for these different comparisons. The DPO point estimate and order consistency describe only the saved DPO comparison.

\begin{figure}[!htbp]
\centering
\includegraphics[width=\textwidth]{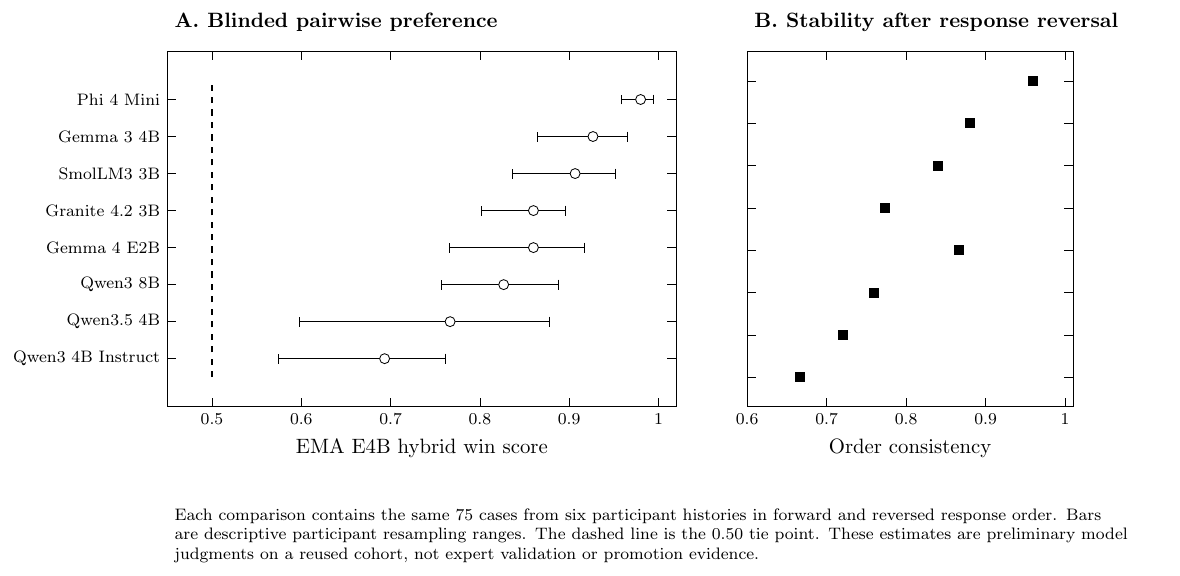}
\caption{Preliminary GPT 5.6 Luna comparisons between \system{} and each external instruction model. Panel A shows the hybrid win score and descriptive participant resampling range. Panel B shows agreement after response order was reversed. Each comparison uses the same 75 cases from six participant histories.}
\label{fig:luna-benchmark}
\end{figure}

The external model judge analysis produced a different pattern. \system{} received a win score above 0.50 against every external model. The lowest value was 0.693 against Qwen3 4B Instruct, and the highest was 0.980 against Phi 4 Mini. Order consistency ranged from 0.667 to 0.960. Figure~\ref{fig:luna-benchmark} reports all eight comparisons. This pattern may reflect task adaptation, contract conformity, explanation style, or the fitted head. The study did not run a judge sensitivity restricted to cases where both responses passed every contract. The result remains preliminary because the cohort contains only six participant histories and had been used before this judge analysis.

\subsection{Judge protocol}

GPT 5.6 Luna at xhigh reasoning was accessed through the Codex SDK on August 26, 2026 for DPO versus the supervised hybrid and on September 2, 2026 for the eight external model comparisons. No sampling temperature or immutable service snapshot was exposed to the study code. The judge used a fixed rubric for forecast, questions, timing, rationale, burden, safety, and overall quality. It saw observable history, the later observed response, and both raw policy and forecast outputs. Model names, training conditions, case identifiers, and participant identifiers were hidden. Candidate position was balanced as 38 A and 37 B assignments, and each case was judged again after response order was reversed. Calls contained at most eight cases for both the E4B and external comparisons. Invalid duplicate identifiers caused rejection and a complete retry rather than repair. 

The DPO judge set contains 75 development cases from eight participant histories; all 22 expert reviewed cases are a subset. The external judge set uses 75 development cases from six of those histories, with seven cases that overlap the completed expert reviews. We used the same external set for each of eight model comparisons. A case receives a candidate win score of one only when both orders favor the candidate, zero only when both favor the comparator, and 0.5 for a consistent tie or an order disagreement. Order consistency is the fraction with the same physical winner after both orders are mapped back to the physical response. This agreement gated score is not the usual fraction of independent pairwise wins: a win in one order and a tie in the other receives 0.5. Rating differences average candidate minus comparator scores over the two orders. Malformed model responses remain unchanged. These model judgments are preliminary and do not replace expert review.

\section{Supplementary Implementation and Analysis}\label{app:implementation}
\subsection{Benchmark checkpoints and engineering measurements}

Table~\ref{tab:checkpoints} records the exact external checkpoint revisions and local runtime measurements. The benchmark used Transformers 5.16.1 and PyTorch 2.7.1 with CUDA 12.8 on an RTX 5090. Qwen3.5 used the native Transformers PyTorch recurrent implementation because the installed optional causal convolution extension did not contain a compatible RTX 5090 kernel image. This substitution changed only the execution kernel, not the checkpoint, prompt, decoding rule, or score.

\begin{table*}[!htbp]
\centering
\caption{Pinned external checkpoints and runtime measurements.}
\label{tab:checkpoints}
\footnotesize
\begin{tabularx}{\textwidth}{lXrr}
\toprule
Checkpoint & Revision & Mean seconds & Peak GiB \\
\midrule
\texttt{HuggingFaceTB/SmolLM3-3B} & \texttt{a07cc9a04f16550a088caea529712d1d335b0ac1} & 13.38 & 6.30 \\
\texttt{ibm-granite/granite-4.2-3b} & \texttt{b7e947307dd2efb3ad3b853b0e8a7e75f8ad4ac2} & 13.42 & 7.40 \\
\texttt{microsoft/Phi-4-mini-instruct} & \texttt{cfbefacb99257ffa30c83adab238a50856ac3083} & 9.19 & 7.93 \\
\texttt{Qwen/Qwen3-4B-Instruct-2507} & \texttt{cdbee75f17c01a7cc42f958dc650907174af0554} & 12.61 & 8.35 \\
\texttt{unsloth/gemma-3-4b-it} & \texttt{bf46152c47f5dd20b896357cb51abc4c03b8ee8c} & 19.25 & 8.90 \\
\texttt{Qwen/Qwen3.5-4B} & \texttt{851bf6e806efd8d0a36b00ddf55e13ccb7b8cd0a} & 18.80 & 9.52 \\
\texttt{unsloth/gemma-4-E2B-it} & \texttt{d36bdd3855c82a6a1a23f7c459b749f5724ae75d} & 23.19 & 10.10 \\
\texttt{Qwen/Qwen3-8B} & \texttt{b968826d9c46dd6066d109eabc6255188de91218} & 11.93 & 16.22 \\
\bottomrule
\end{tabularx}
\end{table*}

\begin{figure}[!htbp]
\centering
\includegraphics[width=\textwidth]{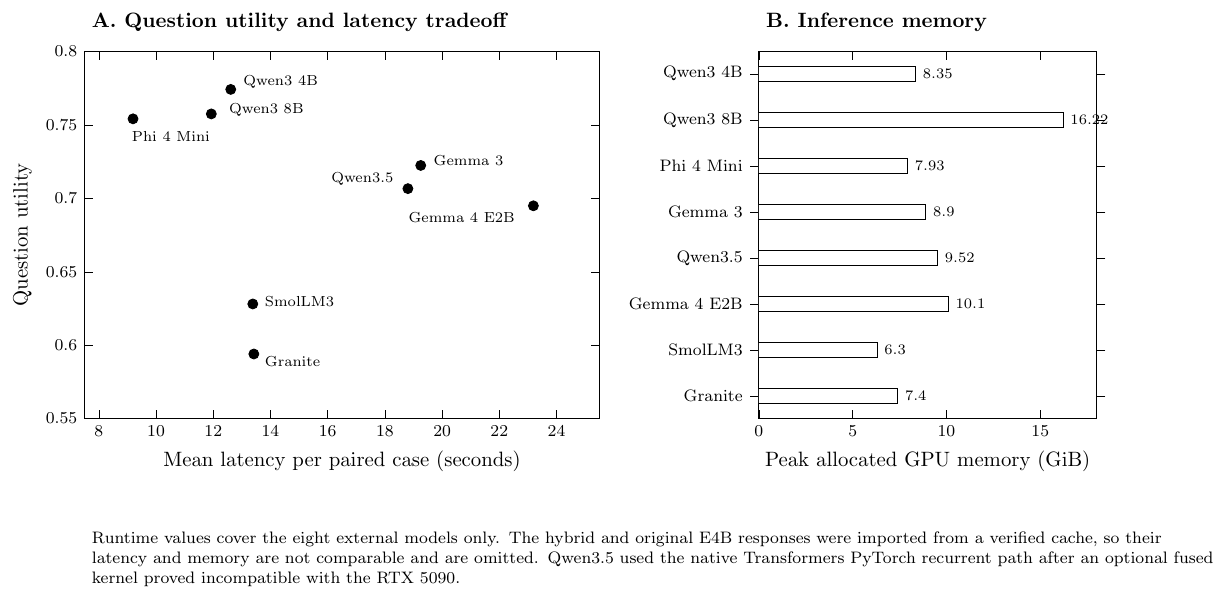}
\caption{Runtime comparison for the eight external checkpoints. Panel A relates question utility to mean paired generation latency. Panel B reports peak allocated GPU memory. Cached E4B references are omitted because comparable runtime measurements are unavailable.}
\label{fig:benchmark-efficiency}
\end{figure}

The external models required 9.19 to 23.19 seconds per paired policy and forecast case in one run per checkpoint. Peak allocated GPU memory ranged from 6.30 to 16.22 GiB. Qwen3 8B used the most memory but did not exceed Qwen3 4B on question utility. Figure~\ref{fig:benchmark-efficiency} reports these descriptive engineering measurements. They are confounded by output length, wrapper failure, model family, and reasoning configuration. No comparable runtime was recorded for \system{}.

\begin{table*}[!htbp]
\centering
\caption{All case benchmark metrics on the fixed 75 case development cohort. The hybrid includes a fitted ridge head that the direct instruction comparators did not receive.}
\label{tab:benchmark}
\small
\setlength{\tabcolsep}{5.8pt}
\begin{threeparttable}
\begin{tabular}{lrrrrrr}
\toprule
System & Policy parse & Exact budget & Language constraint & Question utility & Timing similarity & Forecast wrapper \\
\midrule
\textbf{\system{} hybrid} & \textbf{1.000} & \textbf{1.000} & \textbf{1.000} & \textbf{0.832} & \textbf{0.818} & \textbf{1.000} \\
Qwen3 4B Instruct & 1.000 & 1.000 & 0.907 & 0.774 & 0.390 & 1.000 \\
Original E4B & 1.000 & 1.000 & 0.920 & 0.764 & 0.307 & 1.000 \\
Qwen3 8B & 1.000 & 1.000 & 0.973 & 0.757 & 0.403 & 0.000 \\
Phi 4 Mini & 1.000 & 1.000 & 0.253 & 0.754 & 0.305 & 0.000 \\
Gemma 3 4B & 1.000 & 1.000 & 0.000 & 0.722 & 0.305 & 0.000 \\
Qwen3.5 4B & 1.000 & 1.000 & 1.000 & 0.707 & 0.461 & 1.000 \\
Gemma 4 E2B & 0.987 & 0.987 & 0.947 & 0.695 & 0.419 & 0.000 \\
SmolLM3 3B & 0.973 & 0.973 & 0.453 & 0.628 & 0.310 & 0.000 \\
Granite 4.2 3B & 0.800 & 0.800 & 0.080 & 0.594 & 0.257 & 0.000 \\
\bottomrule
\end{tabular}
\begin{tablenotes}\footnotesize
\item The evaluated model is the ridge plus E4B hybrid assessed in the formative expert review, without DPO. Language constraint is a lexical and structural research screen, not a clinical safety measure. Forecast wrapper requires the exact answer wrapper. Failures remain in every denominator.
\end{tablenotes}
\end{threeparttable}
\end{table*}
\begin{figure}[!htbp]
\centering
\includegraphics[width=.94\textwidth]{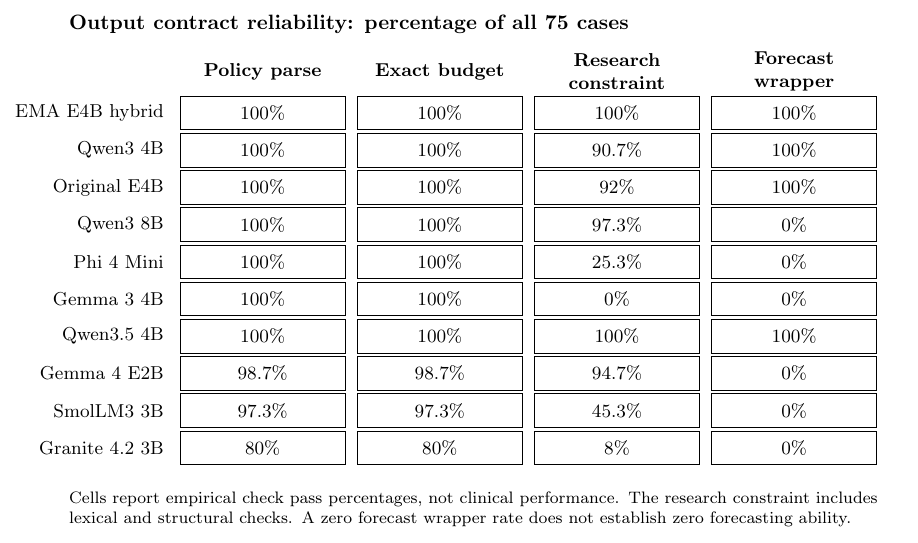}
\caption{Output contract reliability for all ten systems. Policy parse and exact budget do not guarantee the research language constraint or exact forecast wrapper. Every rate uses all 75 cases. Wrapper failures must not be read as forecasting failures.}
\label{fig:benchmark-contract}
\end{figure}
\subsection{Subsequent preference and policy experiments}

Twenty of the 22 expert reviews yielded usable chosen and rejected pairs. A DPO run used rank 16 LoRA, $\beta=0.1$, learning rate $5\times10^{-6}$, and 30 steps. A participant separated experiment trained on 15 pairs from six source participants and evaluated five pairs from two other source participants. All tested checkpoints kept parse, budget, and research language and action constraint rates at 1.0, and their observable actions were unchanged. This analysis does not determine whether explanation quality or other response characteristics changed. The checkpoints were not selected for replacement of the evaluated hybrid.

The separate policy head v2 experiment produced a question utility gain of 0.0100 on 470 one time test cases (case bootstrap interval $[0.0038,0.0162]$). Timing similarity gain was $-0.0017$ ($[-0.0221,0.0186]$), so the analysis cannot resolve effects at the 0.02 gate scale. Five fold participant out of fold estimates were 0.0148 for question utility and 0.0180 for timing similarity. The timing sign differs between the single test and fold estimates. The question estimate remained below the 0.02 gate. This alternative head is not part of the evaluated hybrid.

The 470 case analysis excludes zero based test rows 0, 8, \ldots, 472 and retains all 11 test participants. Its stored rule attributes this exclusion to earlier systematic evaluation. The 60 case language diagnostic in the present paper instead uses development data; the available record does not identify the earlier test consumption event more precisely. The exclusion is preserved rather than retrospectively changed.

\subsection{Participant sensitivity}
An exploratory reanalysis of the 75 case hybrid versus head comparison gives an equal participant mean question difference of $-0.0116$. Omitting each of the six participants in turn gives case weighted differences from $-0.0243$ to $-0.0076$. Timing differences remain zero for every participant. The original case weighted estimates are unchanged. These checks reuse existing responses and are not a new validation cohort.

\subsection{Item mapping and numerical implementation}\label{app:items}
\begin{table}[!htbp]
\centering\small
\caption{Full item to construct mapping used by the code. These are computational groups, not newly validated latent factors.}
\begin{tabularx}{\textwidth}{lXl}
\toprule
Items & Readable measures & Code construct\\
\midrule
Q1, Q2 & Difficulty relaxing; irritability & Stress\\
Q3, Q4 & Worry; nervousness or anxiety & Anxiety\\
Q5, Q6 & Nothing to look forward to; lack of positive feeling & Depression\\
Q9, Q10 & Loneliness; anger & Mood\\
Q7, Q8, Q17 & Fatigue; hunger; coronavirus health concern & Physiological\\
Q11--Q15 & Offline social contact; social media; music; procrastination; outdoors & Behavioral\\
Q16, Q18 & Coronavirus attention; time at home & Context\\
\bottomrule
\end{tabularx}
\end{table}

Trend labels use the last three scheduled rows including the current row, with missing item values omitted. At least two observed values are required. The last minus first difference above 0.25 is increasing, below $-0.25$ is decreasing, and otherwise is stable. With fewer than two values the label is insufficient history. A changing trend means increasing or decreasing. The head's separate numerical endpoint change uses up to three prior rows excluding the current row; it is not a fitted slope or a rate per hour.

Question features are an explicit constant, current value and personal mean divided by five, signed and absolute deviations divided by four, population standard deviation of prior values divided by two, prior last minus first difference divided by four, increasing and decreasing indicators, item index divided by eighteen, and seven construct indicators in alphabetic order. A missing current value becomes zero; a missing personal mean or prior value becomes the current value. There is no learned standardization. The ridge solution is $(X^\top X+\lambda I)^{-1}X^\top y$, which penalizes the explicit intercept too.

Timing features begin with a constant, current risk rank divided by two, positive depression or loneliness elevation divided by three, prompt index modulo four divided by three, the maximum and mean absolute prior endpoint changes divided by four, and the maximum and mean population standard deviations divided by two. The seven nonconstant terms are squared, and three interactions are added: rank times elevation, rank times maximum change, and elevation times maximum change. Missing prior values in these timing summaries become zero. At least two prior scheduled rows exist, but they need not have every item observed.

For question targets, a missing current value becomes zero, while a missing next value or personal mean defaults to that current value. Construct means omit missing items; unavailable construct differences and baseline differences become zero. Pearson correlations use pairwise observed current values of the training cases, not every overlapping history row. Fewer than three paired values or zero variance gives correlation zero. Only absolute correlations of at least 0.45 are stored, rounded to six decimals. No elapsed time normalization enters the target or feature calculation. These are the implementations used for the reported results; they were not changed during revision.

\subsection{Selection and review provenance}
The 200 case generation inventory takes the first 200 development rows: the systematic sampler has stride one because $\lfloor346/200\rfloor=1$. All 200 have valid policy and forecast outputs for both E4B conditions. The 150 case inventory orders each participant's eligible cases by scheduled time, starts near the temporal midpoint, and repeatedly chooses the case farthest in index from those already chosen. Ties favor proximity to the midpoint, then the smaller index. It then cycles through participants in sorted order. The first 75 selected cases form the priority benchmark, and the next 75 form the extension. Reapplying this rule reproduces both sets and their order exactly.

The August 6 expert snapshot contains 22 unique reviews and one saved submission per reviewed case. Its 100 case E4B cache points to detailed paired generations, not the later 150 case inventory. The portal provides a queue and navigation controls; the saved reviews do not establish why particular cases were completed or whether queue order affected completion. The observed 22 case subset is therefore defined by the frozen completed submissions rather than a claimed random sample. All eight external comparisons use precisely the same ordered 75 cases. The DPO judge uses 75 different selected contexts, including the expert contexts, but not the same model contrast.

\subsection{Synthetic worked illustration and schemas}\label{app:synthetic}
\textbf{Synthetic illustration of input and output structure; not an evaluated participant case.} The values below and the illustrative language response are authored examples, not recorded model outputs. They demonstrate what a reader would inspect without releasing a participant trajectory.

Suppose prior depression item values are $(1,1,2)$, loneliness values are $(1,2,3)$, and current values are 2, 2, and 3 for Q5, Q6, and Q9. Other current items equal 1. A hypothetical head proposal selects Q3, Q4, Q5, Q6, Q9, and Q12 at 60 minutes. A preserved response emits that same proposal and explains that the observed loneliness increase motivates review of mood alongside depression and anxiety. A modified response substitutes Q13 for Q9 while retaining 60 minutes; its explanation would need to justify omitting the item with the illustrated increase. This comparison makes action consistency inspectable. Neither hypothetical proposal is claimed to be the fitted head's output, and no empirical utility score is assigned to these authored examples.

The policy schema includes \texttt{selected\_questions}, \texttt{next\_prompt\_delay\_minutes}, \texttt{timing\_label}, \texttt{priority\_level}, \texttt{priority\_construct}, \texttt{expected\_information\_retention}, \texttt{personalization\_basis}, \texttt{expert\_review\_note}, and \texttt{rationale}; the injected proposal also provides \texttt{timing\_urgency}. Expected information retention is a model output, not an independently validated retention estimate. An illustrative forecast response is:
\begin{lstlisting}
<answer>{"next_depression_mean":2.0,"next_loneliness":3.0,
"next_social_media":1.0,"risk_level":"high",
"rationale":"Illustrative persistence estimate from the shown history.",
"uncertainty_note":"Research forecast; requires expert review."}</answer>
\end{lstlisting}
The numbers illustrate the schema only. The high label follows the research threshold on loneliness, not a clinical diagnosis. This forecast concerns the next scheduled record, not an asserted outcome at 60 minutes.

\subsection{What the checks measure}
Policy parsing recovers an answer wrapped JSON object; the item budget check verifies the recovered unique item count. The composite research screen combines payload eligibility, timing label consistency, a structural delay range of 15 to 1,440 minutes, and lexical restrictions. It is not an independent validation of every field in the requested schema. Membership in the five requested delays is a separate property and is not guaranteed by that composite alone. Forecast wrapper conformity detects the exact answer wrapped JSON format, not predictive accuracy. Successful observations in a finite sample do not establish a population reliability of 0.99.

\bibliographystyle{plainnat}
\bibliography{references}

@article{shiffman2008ecological,
  title={Ecological momentary assessment},
  author={Shiffman, Saul and Stone, Arthur A and Hufford, Michael R},
  journal={Annu. Rev. Clin. Psychol.},
  volume={4},
  number={1},
  pages={1--32},
  year={2008},
  publisher={Annual Reviews}
}

@article{williams2021compliance,
  title={Compliance with mobile ecological momentary assessment of self-reported health-related behaviors and psychological constructs in adults: systematic review and meta-analysis},
  author={Williams, Marie T and Lewthwaite, Hayley and Fraysse, Fran{\c{c}}ois and Gajewska, Alexandra and Ignatavicius, Jordan and Ferrar, Katia},
  journal={Journal of medical Internet research},
  volume={23},
  number={3},
  pages={e17023},
  year={2021},
  publisher={JMIR Publications Inc., Toronto, Canada}
}

@article{wrzus2023ecological,
  title={Ecological momentary assessment: A meta-analysis on designs, samples, and compliance across research fields},
  author={Wrzus, Cornelia and Neubauer, Andreas B},
  journal={Assessment},
  volume={30},
  number={3},
  pages={825--846},
  year={2023},
  publisher={Sage Publications Sage CA: Los Angeles, CA}
}

@article{fried2022mental,
  title={Mental health and social contact during the COVID-19 pandemic: An ecological momentary assessment study},
  author={Fried, Eiko I and Papanikolaou, Faidra and Epskamp, Sacha},
  journal={Clinical Psychological Science},
  volume={10},
  number={2},
  pages={340--354},
  year={2022},
  publisher={Sage Publications Sage CA: Los Angeles, CA}
}

@article{speyer2024investigating,
  title={Investigating moderation effects at the within-person level using intensive longitudinal data: A two-level dynamic structural equation modelling approach in Mplus},
  author={Speyer, Lydia Gabriela and Murray, Aja Louise and Kievit, Rogier},
  journal={Multivariate Behavioral Research},
  volume={59},
  number={3},
  pages={620--637},
  year={2024},
  publisher={Taylor \& Francis}
}

@article{nahum2016just,
  title={Just-in-time adaptive interventions (JITAIs) in mobile health: key components and design principles for ongoing health behavior support},
  author={Nahum-Shani, Inbal and Smith, Shawna N and Spring, Bonnie J and Collins, Linda M and Witkiewitz, Katie and Tewari, Ambuj and Murphy, Susan A},
  journal={Annals of behavioral medicine},
  pages={1--17},
  year={2016},
  publisher={Springer}
}

@article{klasnja2015microrandomized,
  title={Microrandomized trials: An experimental design for developing just-in-time adaptive interventions.},
  author={Klasnja, Predrag and Hekler, Eric B and Shiffman, Saul and Boruvka, Audrey and Almirall, Daniel and Tewari, Ambuj and Murphy, Susan A},
  journal={Health psychology},
  volume={34},
  number={S},
  pages={1220},
  year={2015},
  publisher={American Psychological Association}
}

@article{schneider2024just,
  title={Just-in-time adaptive ecological momentary assessment (JITA-EMA)},
  author={Schneider, Stefan and Junghaenel, Doerte U and Smyth, Joshua M and Fred Wen, Cheng K and Stone, Arthur A},
  journal={Behavior research methods},
  volume={56},
  number={2},
  pages={765--783},
  year={2024},
  publisher={Springer}
}

@article{li2024ask,
  title={Ask less, learn more: Adapting ecological momentary assessment survey length by modeling question-answer information gain},
  author={Li, Jixin and Ponnada, Aditya and Wang, Wei-Lin and Dunton, Genevieve and Intille, Stephen},
  journal={Proceedings of the ACM on interactive, mobile, wearable and ubiquitous technologies},
  volume={8},
  number={4},
  pages={1--32},
  year={2024},
  publisher={ACM New York, NY, USA}
}

@article{aminikhanghahi2019context,
  title={Context-aware delivery of ecological momentary assessment},
  author={Aminikhanghahi, Samaneh and Schmitter-Edgecombe, Maureen and Cook, Diane J},
  journal={IEEE journal of biomedical and health informatics},
  volume={24},
  number={4},
  pages={1206--1214},
  year={2019},
  publisher={IEEE}
}

@article{vaswani2017attention,
  title={Attention is all you need},
  author={Vaswani, Ashish and Shazeer, Noam and Parmar, Niki and Uszkoreit, Jakob and Jones, Llion and Gomez, Aidan N and Kaiser, {\L}ukasz and Polosukhin, Illia},
  journal={Advances in neural information processing systems},
  volume={30},
  year={2017}
}

@article{brown2020language,
  title={Language models are few-shot learners},
  author={Brown, Tom and Mann, Benjamin and Ryder, Nick and Subbiah, Melanie and Kaplan, Jared D and Dhariwal, Prafulla and Neelakantan, Arvind and Shyam, Pranav and Sastry, Girish and Askell, Amanda and others},
  journal={Advances in neural information processing systems},
  volume={33},
  pages={1877--1901},
  year={2020}
}

@article{ouyang2022training,
  title={Training language models to follow instructions with human feedback},
  author={Ouyang, Long and Wu, Jeffrey and Jiang, Xu and Almeida, Diogo and Wainwright, Carroll and Mishkin, Pamela and Zhang, Chong and Agarwal, Sandhini and Slama, Katarina and Ray, Alex and others},
  journal={Advances in neural information processing systems},
  volume={35},
  pages={27730--27744},
  year={2022}
}

@article{hu2021lora,
  title={Lora: Low-rank adaptation of large language models},
  author={Hu, Edward J and Shen, Yelong and Wallis, Phillip and Allen-Zhu, Zeyuan and Li, Yuanzhi and Wang, Shean and Wang, Lu and Chen, Weizhu},
  journal={arXiv preprint arXiv:2106.09685},
  year={2021}
}

@article{zheng2023judging,
  title={Judging llm-as-a-judge with mt-bench and chatbot arena},
  author={Zheng, Lianmin and Chiang, Wei-Lin and Sheng, Ying and Zhuang, Siyuan and Wu, Zhanghao and Zhuang, Yonghao and Lin, Zi and Li, Zhuohan and Li, Dacheng and Xing, Eric and others},
  journal={Advances in neural information processing systems},
  volume={36},
  pages={46595--46623},
  year={2023}
}

@article{team2026gemma,
  title={Gemma 4 technical report},
  author={Team, Gemma and Abd, Sherif El and Aggarwal, Vaibhav and Algayres, Robin and Andreev, Alek and Bachem, Olivier and Ballantyne, Ian and Brick, Cormac and C{\u{a}}rbune, Victor and Casbon, Michelle and others},
  journal={arXiv preprint arXiv:2607.02770},
  year={2026}
}

@article{onnela2016harnessing,
  title={Harnessing smartphone-based digital phenotyping to enhance behavioral and mental health},
  author={Onnela, Jukka-Pekka and Rauch, Scott L},
  journal={Neuropsychopharmacology},
  volume={41},
  number={7},
  pages={1691--1696},
  year={2016},
  publisher={Nature Publishing Group}
}

@article{bommasani2021opportunities,
  title={On the opportunities and risks of foundation models},
  author={Bommasani, Rishi and Hudson, Drew A and Adeli, Ehsan and Altman, Russ and Arora, Simran and von Arx, Sydney and Bernstein, Michael S and Bohg, Jeannette and Bosselut, Antoine and Brunskill, Emma and others},
  journal={arXiv preprint arXiv:2108.07258},
  year={2021}
}

@inproceedings{van2025survey,
  title={A survey on small language models},
  author={Van Nguyen, Chien and Shen, Xuan and Aponte, Ryan and Xia, Yu and Basu, Samyadeep and Hu, Zhengmian and Chen, Jian and Parmar, Mihir and Kunapuli, Sasidhar and Barrow, Joe and others},
  booktitle={Proceedings of the 15th International Conference on Recent Advances in Natural Language Processing-Natural Language Processing in the Generative AI Era},
  pages={807--821},
  year={2025}
}

@article{kjell2024beyond,
  title={Beyond rating scales: With targeted evaluation, large language models are poised for psychological assessment},
  author={Kjell, Oscar NE and Kjell, Katarina and Schwartz, H Andrew},
  journal={Psychiatry Research},
  volume={333},
  pages={115667},
  year={2024},
  publisher={Elsevier}
}

@article{guo2024large,
  title={Large language models for mental health applications: systematic review},
  author={Guo, Zhijun and Lai, Alvina and Thygesen, Johan H and Farrington, Joseph and Keen, Thomas and Li, Kezhi},
  journal={JMIR mental health},
  volume={11},
  number={1},
  pages={e57400},
  year={2024},
  publisher={JMIR Publications Inc., Toronto, Canada}
}

@article{rafailov2023direct,
  title={Direct preference optimization: Your language model is secretly a reward model},
  author={Rafailov, Rafael and Sharma, Archit and Mitchell, Eric and Manning, Christopher D and Ermon, Stefano and Finn, Chelsea},
  journal={Advances in neural information processing systems},
  volume={36},
  pages={53728--53741},
  year={2023}
}

@article{shao2024deepseekmath,
  title={Deepseekmath: Pushing the limits of mathematical reasoning in open language models},
  author={Shao, Zhihong and Wang, Peiyi and Zhu, Qihao and Xu, Runxin and Song, Junxiao and Bi, Xiao and Zhang, Haowei and Zhang, Mingchuan and Li, YK and Wu, Yang and others},
  journal={arXiv preprint arXiv:2402.03300},
  year={2024}
}

@article{monacelli2023adaptive,
  title={Adaptive data collection for intraindividual studies affected by adherence},
  author={Monacelli, Greta and Zhang, Lili and Schlee, Winfried and Langguth, Berthold and Ward, Tom{\'a}s E and Murphy, Thomas B},
  journal={Biometrical Journal},
  volume={65},
  number={7},
  pages={2200203},
  year={2023},
  publisher={Wiley Online Library}
}

\end{document}